\documentclass{vgtc}                          

\ifpdf
  \pdfoutput=1\relax                   
  \usepackage{graphicx}                
  \DeclareGraphicsExtensions{.pdf,.png,.jpg,.jpeg} 
\else
  \usepackage{graphicx}                
  \DeclareGraphicsExtensions{.eps}     
\fi%

\graphicspath{{figures/}{pictures/}{images/}{./}} 
\usepackage{enumitem}
\usepackage{microtype}
\usepackage{siunitx}
\usepackage{authblk}
\usepackage{setspace}

\PassOptionsToPackage{warn}{textcomp}  
\usepackage{textcomp}                  
\usepackage{mathptmx}                  
\usepackage{times}                     
\usepackage{cite}                      
\usepackage{tabu}                      
\usepackage{booktabs}                 

\onlineid{1302}

\vgtccategory{Research}

\vgtcinsertpkg

\makeatletter
\renewcommand\AB@affilsepx{\hspace{1cm}\protect\Affilfont}
\makeatother

\title{Effects of Virtual Room Size and Objects on \\ Relative Translation Gain Thresholds in Redirected Walking}

\author[1]{Dooyoung Kim\thanks{e-mail: dooyoung.kim@kaist.ac.kr}}
\author[2]{Jinwook Kim\thanks{e-mail: jinwook.kim31@kaist.ac.kr}}
\author[1]{Jae-eun Shin\thanks{e-mail: jaeeunshin@kaist.ac.kr}}
\author[1]{Boram Yoon\thanks{e-mail: boram.yoon1206@kaist.ac.kr}}
\author[2]{Jeongmi Lee\thanks{e-mail: jeongmi@kaist.ac.kr}}
\author[1,3]{Woontack Woo\thanks{Corresponding author. e-mail: wwoo@kaist.ac.kr}}
\affil[1]{\small KAIST UVR Lab.}
\affil[2]{\small KAIST Visual Cognition Lab.}
\affil[3]{\small KAIST KI-ITC ARRC}

\teaser{
  \includegraphics[width=\textwidth]{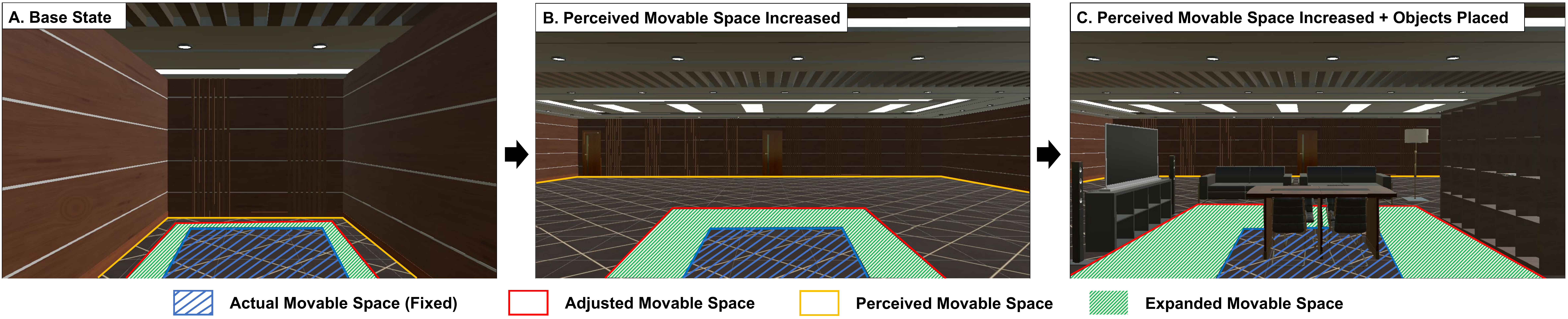}
  \caption{Front views of three different virtual spaces generated for the same movable space in the real world. Compared with (A) the base state, the adjusted movable space to which relative translation gains are applied can be increased when (B) the perceived movable space is larger than the adjusted movable space and (C) objects are placed.}
  \label{fig:teaser}
}

\abstract{
This paper investigates how the size of virtual space and objects within it affect the threshold range of relative translation gains, a Redirected Walking (RDW) technique that scales the user's movement in virtual space in different ratios for the width and depth. While previous studies assert that a virtual room's size affects relative translation gain thresholds on account of the virtual horizon's location, additional research is needed to explore this assumption through a structured approach to visual perception in Virtual Reality (VR). We estimate the relative translation gain thresholds in six spatial conditions configured by three room sizes and the presence of virtual objects (3 $\times$ 2), which were set according to differing Angles of Declination (AoDs) between eye-gaze and the forward-gaze. Results show that both size and virtual objects significantly affect the threshold range, it being greater in the large-sized condition and furnished condition. This indicates that the effect of relative translation gains can be further increased by constructing a perceived virtual movable space that is even larger than the adjusted virtual movable space and placing objects in it. Our study can be applied to adjust virtual spaces in synchronizing heterogeneous spaces without coordinate distortion where real and virtual objects can be leveraged to create realistic mutual spaces.
} 


\CCScatlist{
  \CCScatTwelve{Human-centered computing}{Human Computer Interaction (HCI)}{Interaction paradigms}{Virtual reality};
}

\begin{document}


\firstsection{Introduction}

\maketitle

Redirected Walking (RDW) is a locomotion method for Virtual Reality (VR) based on natural walking. It allows movement in a virtual space larger than the physical space a user is in, and can be used to synchronize a virtual space and a physical space that have different spatial configurations. Of the two main approaches to RDW--adjusting rotation gains or translation gains--the majority of studies on RDW have been focused on the former than the latter on account of the fact that applying rotation gains can modify the virtual space to a greater degree than applying translation gains. However, RDW based on rotation gains often requires instant repositioning to avoid intermittent collisions with obstacles in the physical world, which breaks the sense of immersion. Moreover, whereas the coordinate systems of the real space and virtual space are prone to error, translation gain modification methods can generate a unified coordinate system for both spaces without collision. Therefore, translation gain-based RDW is more appropriate in situations where heterogeneous spaces are converged into a single space for users to share an immersive experience. To increase the effect of modifying a VR client's remote space for optimized mutual space generation, the concept of relative translation gains was proposed to scale the user's movement in a virtual space in different ratios for the width and depth~\cite{kim2021adjusting}.

In order to apply relative translation gains for RDW, it is necessary to estimate the user's threshold range, which refers to the extent to which a user's walking speed can be changed to adjust the virtual movable space without the user noticing the difference in the distance traveled. Previous studies have shown that RDW thresholds are affected by visual cognitive elements such as visual composition of Virtual Environments (VEs) or distractors~\cite{williams2019estimation,nguyen2020effect,kruse2018can}. In the case of relative translation gains, Kim et al.~\cite{kim2021adjusting} found that the threshold range is greater in a large virtual space than in a small one and inferred that the virtual horizon, located higher in the larger space, may have caused users to become more insensitive to changes in their in walking speed therein~\cite{zhao2019exploring}. 
However, they did not present any quantitative data regarding how the users actually perceived the virtual horizon to support this assumption. Although the user's spatial perception of VEs is essential in estimating the RDW threshold range, research focused on the relationship between these two factors has been lacking.

In this study, we explore how the size dimensions and presence of objects configuring a virtual space affect the threshold range of relative translation gains. Based on Kim et al.'s~\cite{kim2021adjusting} study, we set up six experimental conditions by combining three different room sizes (Large, Medium, Small) with the presence of objects (Empty, Furnished). The size conditions were set to disparate the location of virtual horizon and the arrangement of objects were set to affects the Angle of Declination (AoD) between eye-gaze and the forward-gaze. We employed a mixed-subject method where three size conditions were conducted within-subject and the presence of objects proceeded with between-subject. Based on the user's pseudo-Two-Alternative Forced-Choice (pseudo-2AFC) responses on how they felt about their perceived speed in VR, we estimated the threshold range by fitting it to the standard logistic psychometric function. In addition, we measured the actual distribution of AoDs among users for all conditions to analyze the acquired data with regard to the threshold range.

Our statistical results show that both virtual room size and the presence of objects significantly affect the threshold range: The threshold range was biased to be greater in Large condition than in the Small and Medium conditions, and the Furnished condition increased the threshold range to a greater degree than the Empty condition. The distributions of AoD in empty rooms show that users tended to focus their gaze on the virtual horizon for all sizes, which indicates that the size of the room indeed affected the user's visual perception of the space and ultimately the threshold range. On the other hand, the tendency was not apparent in the furnished rooms, where objects functioned more as distractors that contributed to a relative insensitivity to changes in the user's walking speed rather than as blockages to the virtual horizon that decreased the AoD.

Based on our findings, we suggest two design implications to generate virtual spaces where the movable space can be increased to a greater degree with RDW through relative translation gains. First, the perceived movable space in VR should be larger than the adjusted movable space, as illustrated in Figure~\ref{fig:teaser}(B). Second, objects should be placed in the virtual space to increase the relative translation gains further and expand the adjusted movable space, as shown in Figure~\ref{fig:teaser}(C). Our study findings can be applied to create realistic mutual spaces for remote collaboration by adjusting heterogeneous VR clients' spaces with relative translation gains and utilizing the presence of real and virtual objects.

\section{Background}

\subsection{Redirected Walking Techniques}

Various Redirected Walking (RDW) techniques have been introduced to allow users to move around in a more expansive virtual space beyond the limits of physical space, enabling immersive VR experiences~\cite{razzaque2005redirected, langbehn2017application, nilsson2018natural, bozgeyikli2019locomotion}. Some applied curvature gains to steer users to the center of the room or a specific target point~\cite{hodgson2013comparing,chen2018redirected}. There were also RDW techniques with translation gain to move users quickly in large-scale virtual spaces~\cite{interrante2007seven} and analyzed the strength and weakness of translation gain-based locomotion methods according to the location of the eye-level and the virtual space's scale~\cite{abtahi2019m}. RDW methods have also been proposed to construct virtual spaces during a scan and understand the user's physical space~\cite{nescher2016simultaneous}. Dong et al.~\cite{dong2017smooth} introduced Smooth Assembly Mapping (SAM), which decomposes large VE to smaller local patches and mapping together into a real workspace. For multi-user scenarios, Azmandian et al.~\cite{azmandian2017evaluation} explored immersive VR experiences for redirected walking for two users in the same physically tracked space, and Lee et al.~\cite{lee2020optimal} used reinforcement learning in a multi-user environment with heterogeneous physical space.

In addition, Bachmann et al.~\cite{bachmann2019multi} used artificial potential fields to apply RDW and reset algorithms that consider physical obstacles and users. More recently, Williams et al.~\cite{williams2021arc} proposed an alignment-based redirection controller called ARC, displaying the state-of-art performance of real-time, space-adaptive redirection technique that enables free movement. Such redirection techniques using rotational gains~\cite{messinger2019effects,li2019real} were effective in expanding explorable VEs. However, they are prone to collision with the physical environment in free walking scenarios. While RDW using rotation gains is beneficial to increase the size of the virtual movable space, RDW based on translation gains is advantageous in converging spaces with dissimilar configurations into space with unified coordinate system. Although translation gain and rotational gain were utilized together to link spatial information between real and virtual spaces in some cases~\cite{strauss2020steering,williams2021arc}, research focused on translation gains and their potential as an RDW method for mutual space generation has been scarce.

\subsection{Threshold Estimation and Perception}

In order to leverage RDW, it is essential to obtain a threshold range within which users will not be aware of the change in their walking speed. Previous redirection controllers have mostly focused on increasing the adjusted movable area in virtual space or minimizing the number of instance redirections, but did not consider the acceptable range of gains applied to RDW. Estimating this threshold range is necessary because when the user recognizes the difference between their real and virtual movements during RDW, their cognitive performance is affected~\cite{bruder2015cognitive, rietzler2018rethinking, sakono2021redirected}. Steinicke et al.~\cite{steinicke2009estimation} first proposed a RDW threshold estimating method based on measuring the probabilities of users' responses regarding their perceived movement in VEs. Following this study, several studies revisited the concept of threshold range with current VR HMD devices~\cite{grechkin2016revisiting, zhang2018detection}, and some others showed that the threshold range for curvature gains could be increased with repeated RDW trials or continuous walking~\cite{bolling2019shrinking, zhang2014human}, implying that more substantial curvature gain values can be obtained and applied when users grow more familiar with RDW in VE.

Regarding how user perception of VEs affects threshold range values, various cognitive factors were explored, such as the sense of embodiment and the visual composition of VEs~\cite{nguyen2020effect,kruse2018can,neth2012velocity, paludan2016disguising}. Other studies attempted to explore how the size and layout of physical tracked space influence the performance of redirected walking controllers~\cite{azmandian2015physical,messinger2019effects}. In terms of visual perception in VR, previous studies stated users' tend to underestimate virtual walking speeds~\cite{banton2005perception, bruder2011tuning, caramenti2018matching, ash2013vection} and the presence of distractors in VEs bias the threshold range to be larger~\cite{williams2019estimation, gao2020visual,ciumedean2020mission, cools2019investigating}. While Nguyen et al.~\cite{nguyen2018effect} reported that a room's dimension does not significantly affect curvature gain thresholds, Kim et al.~\cite{kim2021adjusting} stated that relative translation gain thresholds differ according to the VE size and referred to Messing et al.~\cite{messing2005distance} in assuming that this may be induced by the location of the virtual horizon. While these studies addressed the relationship between visual perception and threshold range, none of them utilized quantitative data for more detailed and precise analyses.

\subsection{Relative Translation Gains}

Relative translation gains is an extended concept of translation gain that enables spatial deformation without coordinate distortion and enables more space-adaptive modification than uniform translation. Unlike other studies that investigated curvature gain, which utilizes translation gains and rotational gains together, Kim et al.~\cite{kim2021adjusting} suggested relative translation gains, which refer to a pair of translation gains that are applied to the width and depth axes of a plane, respectively, as a method of adjusting the size of virtual spaces. Following Kim et al.'s~\cite{kim2021adjusting} definition, the ratio of the two translation gains in a single pair of relative translation gains (2D translation ratio ($\alpha_T$)) is represented as follows:

\begin{equation}
    \ \alpha_T = \frac{g_{T,x}}{g_{T,y}}\
    \label{equ:relativetranslation}
\end{equation}

where $g_{T,x}$ is the VR environment’s x-axis translation gain, and $g_{T,y}$ is the VR environment’s y-axis translation gain used as the reference translation gain. A relative translation gain threshold consists of two boundary values: maximum 2D translation ratio and minimum 2D translation ratio. In the range between two values of relative translation gain thresholds, a suitable value $g_{T,x}$ can be selected to modify the VR user’s space to generate an optimal virtual movable space. By applying relative translation gains, the VR user's movable space can be rescaled while sustaining the alignment of the coordinate system between the physical and virtual space. Furthermore, relative translation gains can be utilized to create mutual spaces for immersive remote collaboration, as it does not incur sudden changes to the coordinate system.

Kim et al.~\cite{kim2021adjusting} estimated the threshold range of relative translation gains in two differently sized virtual rooms where the larger room (8 m $\times$ 6 m) was four times bigger than the smaller one (4 m $\times$ 3 m). The two estimated 2D translation ratio values consisting relative translation gain thresholds in the large virtual room were 0.85 (lower) and 1.29 (upper). This indicates that the participants' perceived speed in a virtual environment can be changed from the actual speed to be 15\% slower or 29\% faster without them noticing. In the smaller virtual room, the estimated 2D translation ratios were 0.68 (lower) and 1.16 (upper). 

Moreover, significant differences were found in participants' responses regarding their perceived walking speed between the two size conditions. In reporting these results, Kim et al.~\cite{kim2021adjusting} assumed that they may be a result of differences in the location of the virtual horizon, which are determined by room size based on a previous study that perceived distance in the VR space was biased to be larger when the location of horizon became lower~\cite{messing2005distance}. However, no quantitative data was presented to support the argument that relative transition gains depend on the height of the horizon in virtual space. Therefore, our study aims to address the limitations of Kim et al.'s~\cite{kim2021adjusting} work by adopting a more structured approach in verifying how the user's visual perception in VEs affects threshold ranges for relative translation gains.

\begin{figure}[tb]
 \centering
 \includegraphics[width=\columnwidth]{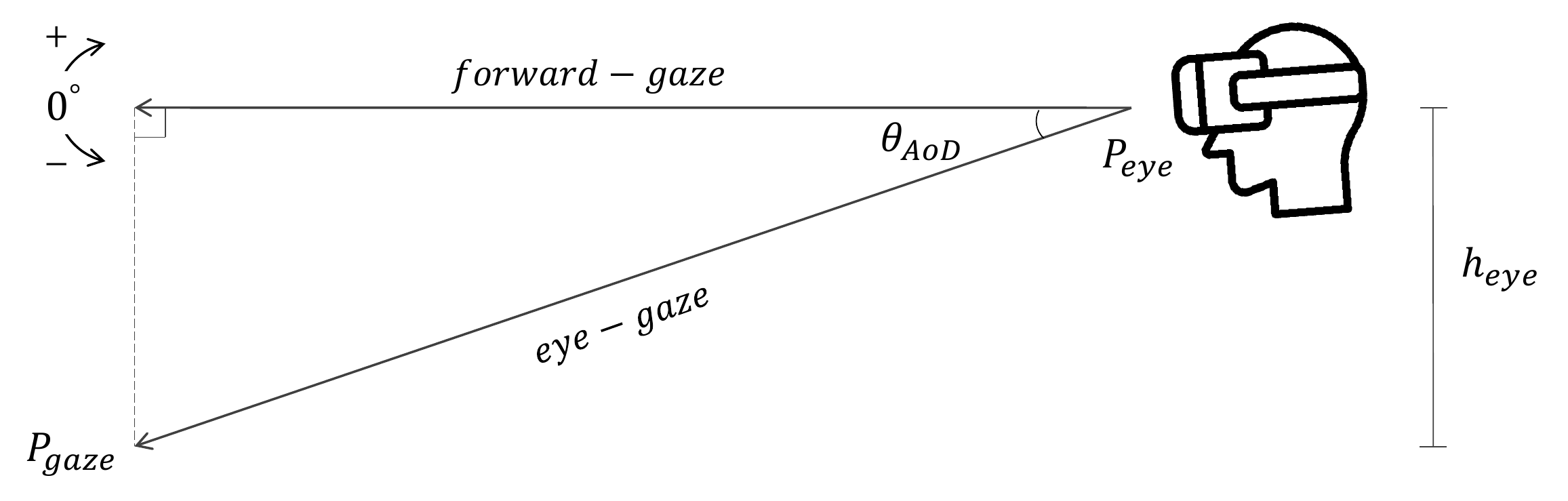}
 \caption{The Angle of Declination (AoD) between the participant's eye-gaze and forward-gaze. Forward-gaze refers to the orthogonal projection vector of the eye-gaze.}
 \label{fig:AoD}
\end{figure}

\section{Method}
\subsection{Research Questions and Hypotheses}

The goal of this study is to investigate the effects of virtual room size and objects on relative translation gain thresholds with more quantitative data and provide recommendations on enlarging the movable area of a virtual space. For this, we decided to use the Angles of Declination (AoD) between the user's eye-gaze and the forward-gaze, as shown in Figure~\ref{fig:AoD}, to configure various spatial conditions. The eye-gaze refers to the vector from the subject's eye position ($P_{eye}$) to the subject's eye-gazed point ($P_{gaze}$), and the forward-gaze refers to the orthogonal projection vector of the eye-gaze. The size of virtual rooms and the location of objects were set according to different locations of virtual horizon formed by the intersection between the room's floor and the wall. We also aim to observe how the users perceive the virtual space in each spatial configuration with a distribution of AoD in each condition. For these goals, we set our research question as follows:

\begin{enumerate}  [label={RQ\arabic*}.,noitemsep]
\item How do virtual room size and object affect the threshold range of relative translation gains?
\item How do virtual room size and object affect the distribution of AoDs?
\end{enumerate}

To answer these questions, we set three room sizes (Large, Medium, Small) and two layouts (Empty, Furnished) for our study conditions. In doing so, we assumed that when the AoD distribution get increases, the user will be more sensitive to changes in their walking speed based on a previous study asserting that the user becomes more sensitive to shortened distances in a VE where the virtual horizon is farther away~\cite{zhao2019exploring}. The Furnished condition was set to shift the AoD distribution to the lower end compared with the Empty condition. In accordance with the rationale behind the study conditions, we derived our hypotheses as below:

\begin{enumerate} [label={H1-\arabic*}.,noitemsep]
	\item The probabilities of ``larger" responses to all relative translation gains will be lower in the Large condition than in the Medium condition.
	\item The probabilities of ``larger" responses to all relative translation gains will be lower in the Large condition than in the Small condition.
	\item The probabilities of ``larger" responses to all relative translation gains will be lower in the Medium condition than in the Small condition.
\end{enumerate}
\begin{enumerate} [label={H2-\arabic*}.,noitemsep]
	\item The probabilities of ``larger" responses to all relative translation gains will be lower in the Large $\times$ Empty condition than in the Large $\times$ Furnished condition.
	\item The probabilities of ``larger" responses to all relative translation gains will be lower in the Medium $\times$ Empty condition than in the Medium $\times$ Furnished condition.
	\item The probabilities of ``larger" responses to all relative translation gains will be lower in the Small $\times$ Empty condition than in the Small $\times$ Furnished condition.
\end{enumerate}

\begin{table}
  \caption{The virtual room size set to a fixed AoD value between the user's ($h_{eye} = 1.5 m$) eye-gaze and the forward-gaze at the center of each room}
  \label{tab:emptyroom}
  \scriptsize%
	\centering%
  \begin{tabu}{%
	*{7}{c}%
	*{2}{r}%
	}
  \toprule
   & Large &  Medium & Small \\
  \midrule
  AoD Setting & $-10^{\circ}$ & $-20^{\circ}$  & $-30^{\circ}$ \\
  Size & 17 m $\times$ 17 m & 8.2 m $\times$ 8.2 m  & 5.2 m $\times$ 5.2 m\\
  \midrule
  \end{tabu}%
\end{table}

\begin{figure}[t]
 \centering
 \includegraphics[width=\columnwidth]{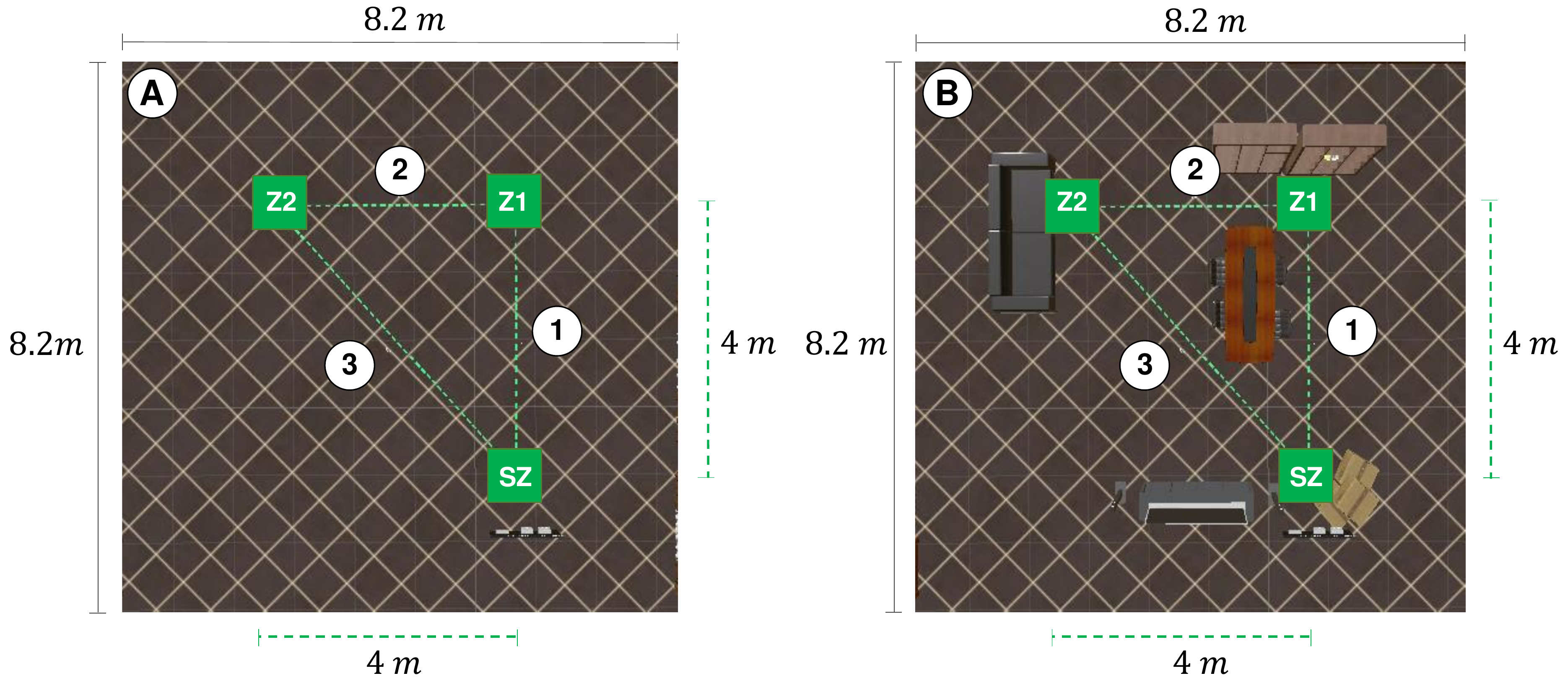}
 \caption{Top-view of the path in (A) Empty and (B) Furnished in Medium size condition. Each number on the floor refers to a sequential walking path that consists the combined paths (SZ = Start Zone, Z1 = Zone 1, Z2 = Zone 2).}
 \label{fig:pathDesign}
\end{figure}

\subsection{Study Design}

\begin{figure*}[t]
 \centering
 \includegraphics[width=\textwidth]{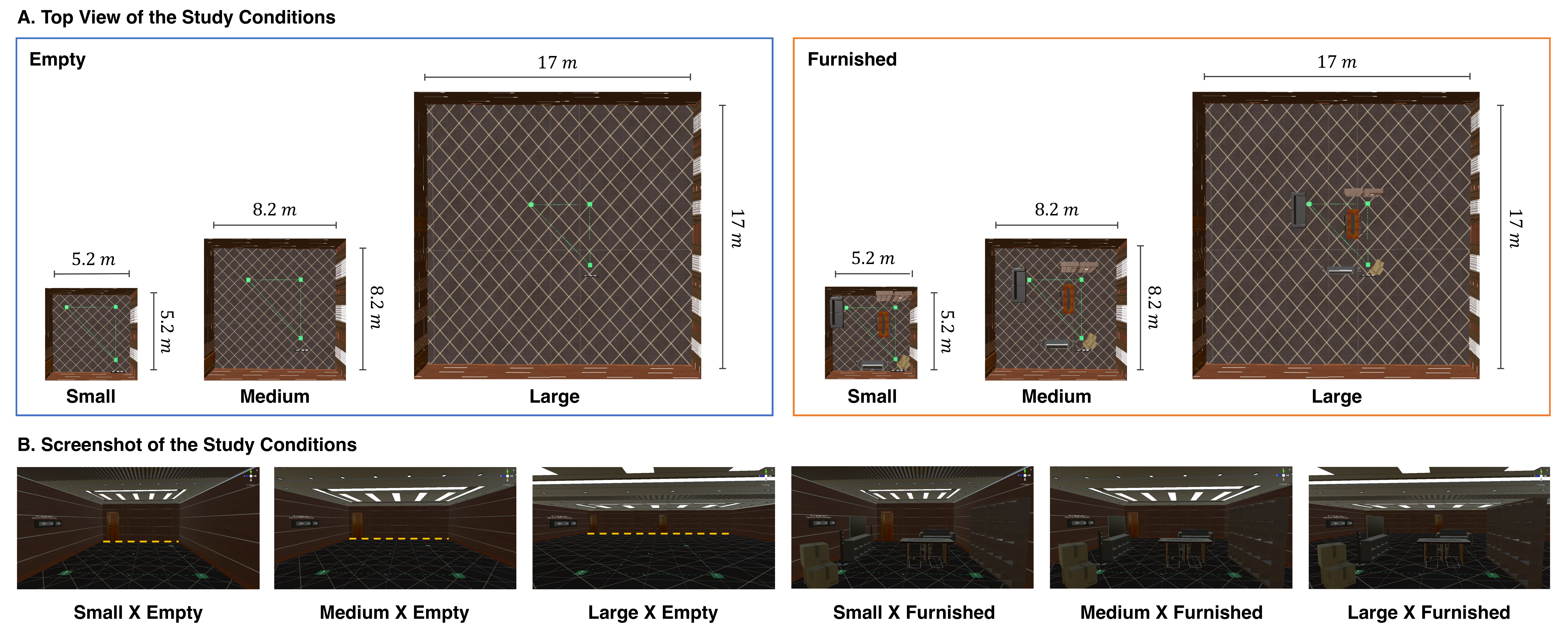}
 \caption{(A) Top-view of the six study conditions. (B) Screenshot of the six study conditions and the yellow dashed lines refer the virtual horizon in each size condition. Combined path and furniture are placed at the center of each condition.}
 \label{fig:conditions}
\end{figure*}

We chose three paths out of five paths used in Kim et al.~\cite{kim2021adjusting}, considering the location of objects in the Furnished condition. Figure~\ref{fig:pathDesign} shows the three paths we used for our study. Subjects repeatedly performed the walking tasks in sequence from path 1 to path 3. Through path 1 and path 2, participants experienced the maximum translation gain and minimum translation gain. On path 3, they experienced the square mean of the maximum and minimum translation gain. Following relative translation gains conditions used for threshold estimation in Kim et al.~\cite{kim2021adjusting}, we fixed a reference translation gain ($g_{T,y}$) to 1.0. Another axis’s translation gain $g_{T,x}$ was set at 0.825, 0.875, or 0.925 to estimate the minimum 2D translation ratio, and $g_{T,x}$ was set to 1.15, 1.2, or 1.25 to estimate the maximum 2D translation ratio. These six relative translation gains were randomly ordered and repeated seven times with the Latin-squared method for within-counterbalancing.

As shown in Figure~\ref{fig:AoD}, the AoD ($\theta_{AoD}$) between eye-gaze and the forward-gaze can be measured from the position of the eye-gazed point ($P_{gaze}$) and the position of the eye ($P_{eye}$). We can compute $\theta_{AoD}$ with the following formula:

\begin{equation}
    \ \theta_{AoD} = \arctan (\frac{y_{gaze} - y_{eye}}{\sqrt{(x_{gaze} - x_{eye})^2 + (z_{gaze} - z_{eye})^2}})
    \label{equ:theta_AoD}
\end{equation}

where the position of the eye-gazed point is $P_{gaze} = (x_{gaxe}, y_{gaze}, z_{gaze})$ and the position of the eye is $P_{eye} = (x_{eye}, y_{eye}, z_{eye})$. $P_{gaze}$ and $P_{eye}$ were measured using HTC Vive Pro Eye, SPanipal SDK, and TobiiXR plugin used in the experiment. To determine the virtual room size, we assumed that a participant whose eyes are 1.5 m above the floor surface ($h_{eye}$) is located at the center of the virtual room. We then deployed a square-shaped virtual conference room with walls located at the virtual horizon set for the preconfigured AoD values. As shown in Table~\ref{tab:emptyroom}, we set the AoD to be $-10^{\circ}$, $-20^{\circ}$, $-30^{\circ}$. In order to induce the following AoD setting, lengths and widths of each VR room were set at 17 m, 8.2 m, 5.2 m. We named each VR room's size condition as Large (17 m $\times$ 17 m), Medium (8.2 m $\times$ 8.2 m), and Small (5.2 m $\times$ 5.2 m).

In the Furnished condition, virtual objects were placed to shorten the distance to the virtual horizon, thereby shifting the AoD distribution to the lower end compared with the Empty condition. In order to do so, we chose a fixed layout that could block the user's eye-gaze from reaching the wall with 5 clusters of virtual furniture items commonly found in a conference room: two bookshelves, a sofa, three boxes, a table with four chairs, and a TV with speakers. To decide the location of each cluster of furniture, we divided them into two groups. The first group's role was to block eye-gaze to reach the virtual horizon made by the wall, and three sets of furniture were placed as Figure~\ref{fig:pathDesign}(B). Two bookshelves were placed at the end of path 1 to prevent user's eye-gaze to reach to the wall while they were walking in zone 1. Similarly, the sofa was located at the end of path 2 and three boxes were placed at the end of path 3. The other two clusters (A table cluster and a TV cluster) were deployed for more realistic conference room construction. We named the virtual room with furniture as the Furnished condition and the empty room as Empty condition.

\begin{figure*}[tb]
 \centering
 \includegraphics[width=\textwidth]{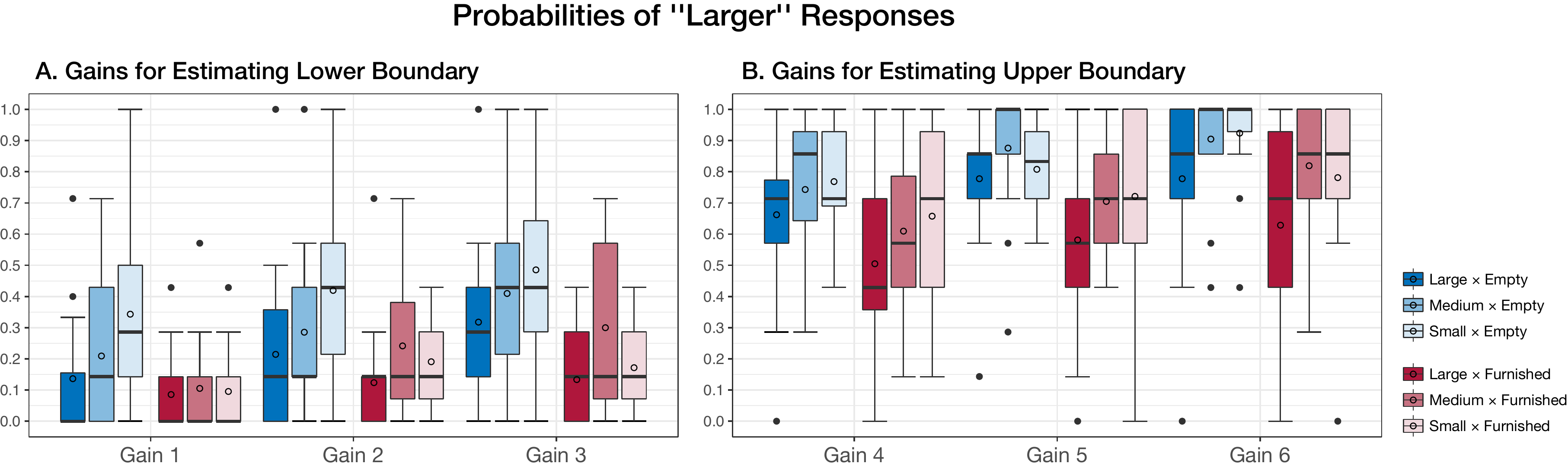}
 \caption{The effects of Size and Virtual Object on the probabilities of ``larger'' responses (o: mean)}
 \label{fig:ARTResult}
\end{figure*}

By combining three room sizes (Large, Medium, Small) with two object layouts (Empty, Furnished), we generated six experimental conditions to estimate relative translation gain thresholds, as illustrated in Figure~\ref{fig:conditions}. To identify the effect of relative translation gains on VR sickness, we added two additional conditions without applying relative translation gains to the study. The VR sickness will increase when users wear VR HMD for a long time, so if the RDW technique affects the increase of VR sickness, this RDW is hard to apply for the actual scenario. In order to see the effect of relative translation gains on VR sickness, we compared post-Simulator Sickness Questionnaire (SSQ)~\cite{kennedy1993simulator} scores with 42 trials without relative translation gains and 42 trials with relative translation gains applied. As Kim et al.~\cite{kim2021adjusting} found that the post-SSQ were higher in a larger virtual space, we chose the Large condition to observe the effect of relative translation gains on VR sickness. We could finalize a total of eight conditions: Small $\times$ Empty, Medium $\times$ Empty, Large $\times$ Empty, Large $\times$ Empty (No Gains), Small $\times$ Furnished, Medium $\times$ Furnished, Large $\times$ Furnished, Large $\times$ Furnished (No Gains).

The study was set up in an empty 6 m $\times$ 6 m physical indoor space with four HTC VIVE base stations (v2.0) installed at the top corners. We used Unity 3D (v2019.3.7f1) and steam VR Plugin (v1.16.10) to implement a virtual conference room environment. Each room was established based on the same conference room prefab, and the combined paths and furniture were located at the center of each room. We employed  SRanipal SDK and TobiiXR Plugin to obtain each participant’s eye-tracking data. We measured the AoD through Equation~\ref{equ:theta_AoD} approximately 36 $\theta_{AoD}$ samples per second. From the accumulated AoD values, we extracted ones obtained when users made a change in their paths and when they were standing at a fixed position in the Start Zone to answer questionnaires.

\subsection{Participants}

\begin{table}
  \caption{Participants' information in two groups}
  \label{tab:participant}
  \scriptsize%
	\centering%
  \begin{tabu}{%
	*{7}{c}%
	*{2}{r}%
	}
  \toprule
   Subjects & Participant Group 1 &   Participant Group 2 \\
  \midrule
  Experimental Condition & Empty & Furnished \\
  Number of Male & 10 & 10  \\
  Number of Female & 6 & 6  \\
  Average Height (Male) & 175.4 cm (SD = 5.2) & 175.1 cm (SD = 7.4) \\
  Average Height (Female) & 163.2 cm (SD = 6.5) & 163.3 cm (SD = 1.9)\\
  \midrule
  \end{tabu}%
\end{table}

The participants were recruited through the local university website and paid \$ 20 in remuneration. Of the 32 participants, 20 identified as male and 12 as female. All participants were at least 18 years old and all had normal vision or corrected to normal vision. The mean age of the participants was 23.81 (SD = 3.86), and the mean interpupillary distance (IPD) of them was 64.17 (SD = 1.89). Most of them had a moderate level of experience in HMD-mediated VR environments: 24 participants had worn VR HMDs up to ten times before and four for more. Only four of them had no prior experience. To reduce the learning effect and fatigue from using VR HMDs, we separated participants into two groups and opted for a mixed-subject method with the existence of virtual objects as the between-subject variable and the size of the VR room as the within-subject variable. As we assume the AoD may affect the threshold range, we balanced the average height between the two groups, as given in Table~\ref{tab:participant}. In addition, we also maintained the gender ratio among the groups, as it also might affect the outcome~\cite{williams2019estimation}.

\subsection{Study Procedure}

\begin{figure}[tb]
 \centering
 \includegraphics[width=\columnwidth]{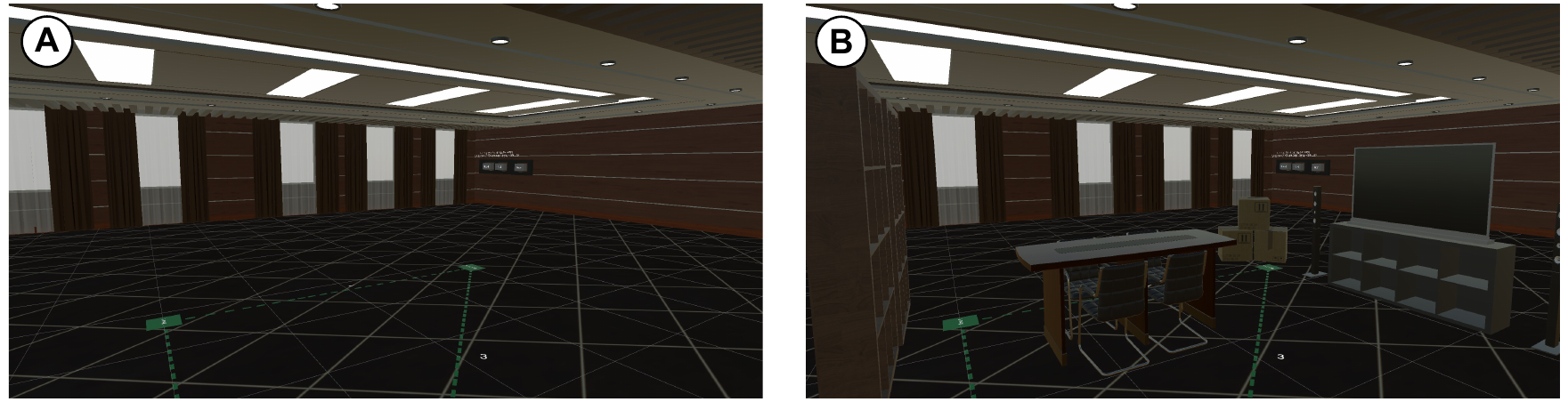}
 \caption{The participant's view through the VR HMD during the experiment. (A) Large $\times$ Empty, (B) Large $\times$ Furnished.}
 \label{fig:Participant}
\end{figure}

The content and procedures of this study were approved by an Institutional Review Board. The first group of participants experienced four Empty conditions, and the other group experienced four Furnished conditions. Figure~\ref{fig:Participant} show the participant's view through the VR HMD during the experiment. Figure~\ref{fig:Participant}(A) is participant's view in Large $\times$ Empty condition where the first group of participants experienced and Figure~\ref{fig:Participant}(B) show the participant's view of Large $\times$ Furnished condition that the second group of participants experienced. They walked along the three given paths in sequential order while experiencing changes to their perceived speed between the maximum and minimum translation gain values. After they walked along three paths, they were asked to answer whether they felt their movements in the virtual environment were “larger” or “smaller” than in the real environment. We used a pseudo-Two-Alternative Forced-Choice (pseudo-2AFC) which is widely used for estimating redirected walking threshold range~\cite{steinicke2009estimation,williams2019estimation,grechkin2016revisiting}. After they answered this questionnaire and pressed the “next” button, they went to the Start Zone and repeated 42 trials (six relative translation gains $\times$ seven repetitions). The order of the four conditions conducted in each participant group was randomized and counterbalanced.

In the study sessions, participants first wore an HTC VIVE Pro Eye with a wireless HTC Vive Pro adapter attached to their heads and held an HTC VIVE Pro controller in their hands to completed an eye-tracking calibration session. They then conducted test trials before the main study trials, which included walking along the path and answering the pseudo-2AFC questionnaire. After the test trial, they took off the VR HMD and filled out the pre-SSQ. Next, they conducted the main study trials. They were permitted to take breaks whenever they wanted and were required to answer the post-SSQ after each condition was completed. After they finished every condition, they were subject to a short semi-structured interview on how the VR room's configuration affected their response to the questionnaire. Each group of participants experienced 168 trials (42 trials $\times$ four experimental conditions) and took approximately 15 minutes for each experimental conditions. The total duration of each study session was approximately two hours.

\begin{figure*}[tb]
 \centering
 \includegraphics[width=\textwidth]{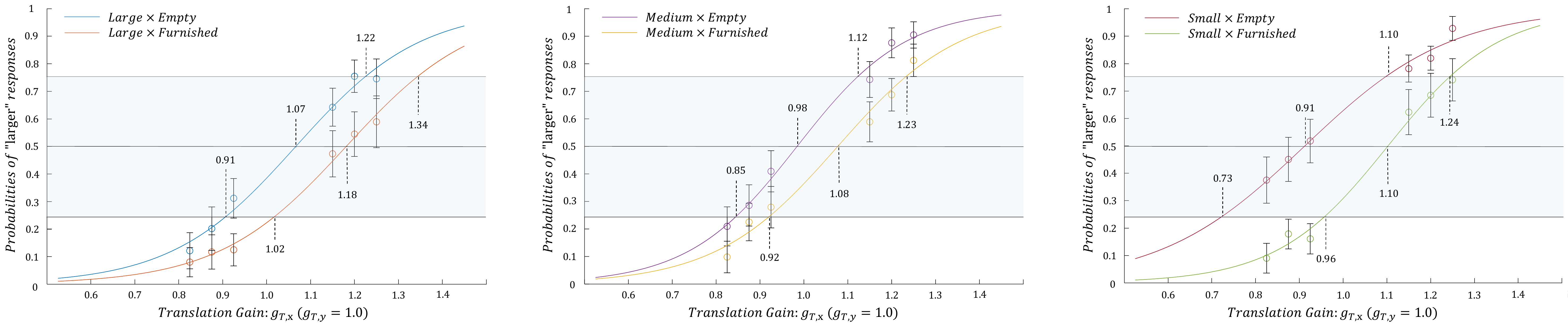}
 \caption{Fitted psychometric functions of mean estimated threshold values in six virtual room conditions}
 \label{fig:threshold}
\end{figure*}

\section{Result}

\subsection{Statistic Result}

We investigated how the size of a virtual space (\textit{Size}) and the placement of objects in it (\textit{Object}) affect the user's relative translation gain thresholds (\textit{Gain}) by comparing the probabilities of ``larger'' responses. The Aligned Rank Transform (ART) for non-parametric factorial ANOVA analysis ($\alpha = 0.05$), proposed by Wobbrock et al. \cite{wobbrock2011aligned}, was applied to conduct a multivariate analysis for the within-subject factor of \textit{Size} (Large, Medium, Small), the between-subject factor of \textit{Object} (Empty and Furnished), and the factor of \textit{Gain} (Gain 1 ~ Gain 6). All pairwise comparisons during the post-hoc analysis were Bonferroni corrected. The six relative translation gain factors represent the following pairs of x-axis and y-axis translation gain values:
Gain 1 = (0.825, 1), Gain 2 = (0.875, 1), Gain 3 = (0.925, 1), Gain 4 = (1.15, 1), Gain 5 = (1.20, 1), and Gain 6 = (1.25, 1). We excluded two participants' data as outliers, one in the Empty condition on account of system malfunction and one in the Furnished condition due to participant error.

We found significant main effects of \textit{Size} ($F(2,490) =$ 16.690, $p<$ .001), \textit{Object} ($F(1,490) =$ 60.984, $p<$ .001), and \textit{Gain} ($F(5,490) =$ 140.894, $p<$ .001). 
The pairwise comparison revealed significant differences between pairs of \textit{Size} conditions: Large and Medium ($p<$ .001), and Large and Small ($p<$ .001). A significant difference between the Medium and Small condition ($p>$ .05) was not found. In the post-hoc analysis, four pairs of \textit{Gain} levels showed no significant differences (Gain 1-Gain 2: $p=$ .109; Gain 2-Gain 3: $p=$ .977; Gain 4-Gain 5: $p=$ .255; and Gain 5-Gain 6: $p=$ .517). On the other hand, all other pairs showed significant differences (all $p<$ .001). Conclusively, we found that the mean value for the probability of ``larger'' responses was significantly lower in the Large room than both the Medium and Small room for all six relative translation gains, as shown in Figure~\ref{fig:ARTResult}. Furthermore, the mean probability value for the Furnished room was also significantly lower than for the Empty room across all gains.

\subsection{Threshold Estimation}

\begin{table}[t]
  \caption{Relative translation gain thresholds according to virtual room configurations}
  \label{tab:threshold}
  \scriptsize%
	\centering%
  \begin{tabu}{%
	*{7}{c}%
	*{2}{r}%
	}
  \toprule
   VE Configuration & $\alpha_{T,lower}$ (25\%) &   PSE (50\%) &   $\alpha_{T,upper}$ (75\%) \\
  \midrule
  Large $\times$ Empty & 0.91 & 1.07  & 1.22 \\
  Medium $\times$ Empty & 0.85 & 0.98  & 1.12 \\
  Small $\times$ Empty & 0.73 & 0.91  & 1.10 \\
  Large $\times$ Furnished & 1.02 & 1.18  & 1.34 \\
  Medium $\times$ Furnished & 0.92 & 1.08  & 1.23 \\
  Small $\times$ Furnished & 0.96 & 1.10  & 1.24 \\
  \midrule
  \end{tabu}%
\end{table}

The results of fitted psychometric functions of mean estimated threshold values in six virtual room conditions are given in Figure~\ref{fig:threshold}. The mean estimated threshold values of participants and Standard Error of the Mean (SEM) are also presented in Figure~\ref{fig:threshold}. The graph's x-axis shows the translation gain for the x-axis ($g_{T,x}$) in the VR room where the reference translation gain ($g_{T,y}$) is fixed at 1.0. The graph's y-axis shows the probabilities of participants’ responses to the question, ``Was the virtual movement larger or smaller than the physical movement?" We used a standard logistic psychometric function to fit the data as follows:

\begin{equation}
    \ f(x)=\frac{1}{1+e^{ax+b}}\
    \label{equ:pse}
\end{equation}

The Point of Subjective Equality (PSE) of each condition was obtained from the fitted psychometric functions. To detect threshold value (2D translation ratio, $\alpha_T$), we applied the 25\% and 75\% criterion used by Steinicke et al.~\cite{steinicke2009estimation}. Table~\ref{tab:threshold} shows the estimated relative translation gain thresholds of each condition. $\alpha_{T,lower}$ refers to the lower boundary of a threshold range (minimum 2D translation ratio), and $\alpha_{T,upper}$ refers to the upper boundary of a threshold range (maximum 2D translation ratio), which indicate that the participant's perceived speed of one axis in a VE can be slower or faster between $\alpha_{T,lower}$ and $\alpha_{T,upper}$ without the user noticing the differences. Estimated threshold values increased according to the size increases in Empty conditions. Moreover, these threshold values were higher in the Furnished condition than in the Empty condition for the same room size.

\subsection{AoD Distribution}

\begin{figure}[tb]
 \centering
 \includegraphics[width=\columnwidth]{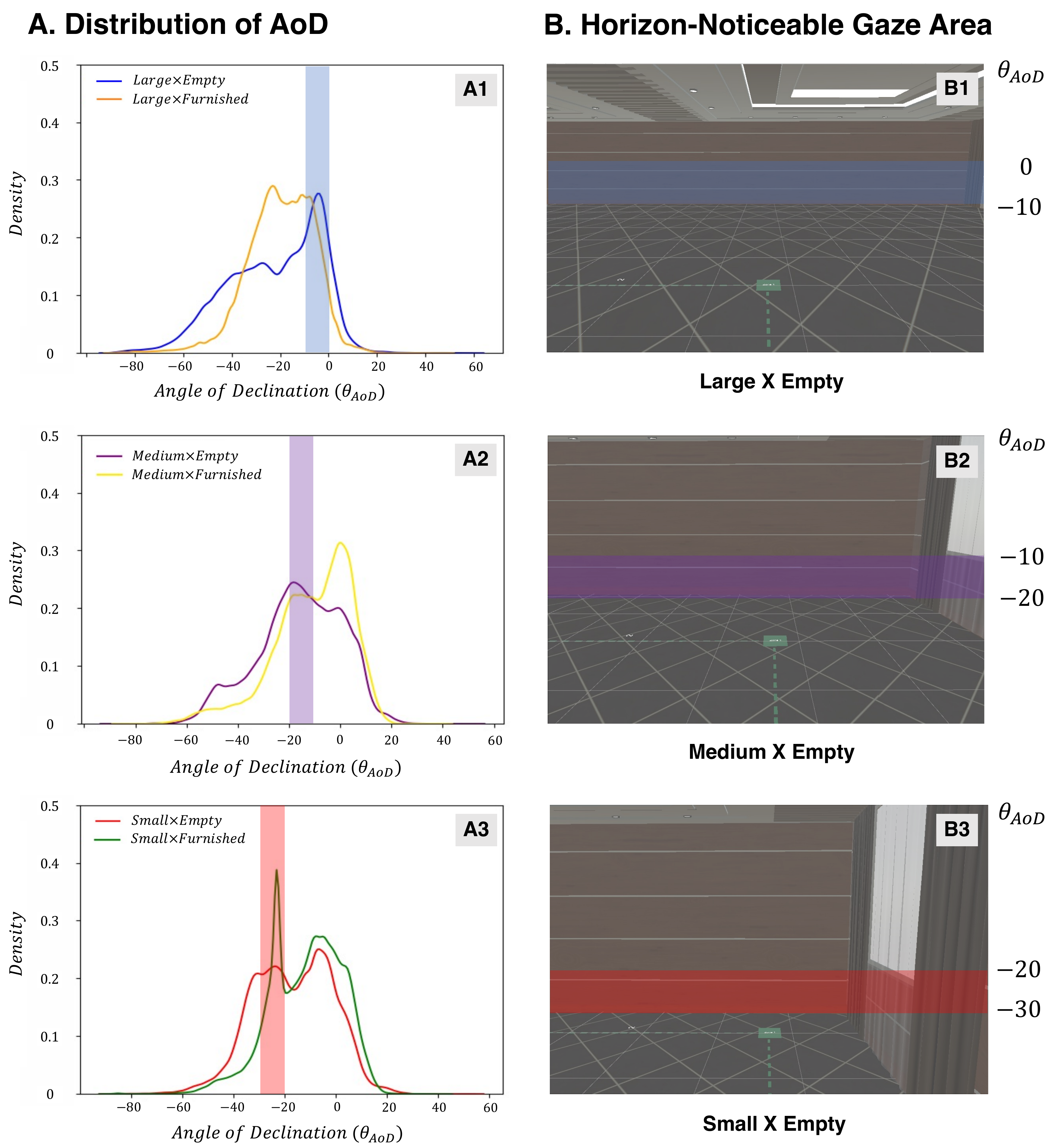}
 \caption{(A) The density distribution of AoD according to VE configurations. (B) The horizon-noticeable gaze area where users can perceive the virtual horizon for (A1) Large, (A2) Medium, and (A3) Small conditions. The highlighted area with color refer the AoD area where the user could perceive the virtual horizon in each size conditions ((A1),(B1) Blue: Large $\times$ Empty, (A2), (B2) Purple: Medium $\times$ Empty, (A3), (B3) Red: Small $\times$ Empty)}
 \label{fig:AoDResult}
\end{figure}

Figure~\ref{fig:AoDResult}(A) shows the distribution of AoD according to each VR room size condition. The x-axis of the graph is the AoD ($\theta_{AoD}$) between the user's eye-gaze and the forward-gaze. $\theta_{AoD} = 0^{\circ}$ means a participant is gazing straight forward when they are walking. When they gaze downward, $\theta_{AoD}$ decreases to a negative degree, and reversely, $\theta_{AoD}$ increases to a positive degree when they gaze upward. The y-axis of the graph shows the density of normalized AoD for each participant. The AoD degree with high density means many participants gaze to that particular AoD. We measured participant's AoD from the position of the foveal area, but humans typically recognize about 10 degrees downward from the location of the foveal area while they are walking~\cite{hill1986preferred}. In the case of Large $\times$ Empty, the average AoD was set to  $-10^{\circ}$ so that the user could notice the virtual horizon in areas ranging from $0^{\circ}$ to $-10^{\circ}$, Medium $\times$ Empty from $-10^{\circ}$ to $-20^{\circ}$, and Small $\times$ Empty from $-20^{\circ}$ to $-30^{\circ}$. 

Based on this, we highlighted the AoD area where users could perceive the room's virtual horizon with the corresponding color in each graph. The blue areas in Figure~\ref{fig:AoDResult}(A1),(B1) refers to the virtual horizon-noticeable area of Large $\times$ Empty, the purple areas in Figure~\ref{fig:AoDResult}(A2),(B2) to those of Medium $\times$ Empty, and the red areas in Figure~\ref{fig:AoDResult}(A3),(B3) to those of Small $\times$ Empty. Figure~\ref{fig:AoDResult}(A1) shows that in the Large condition, the presence of virtual objects biases the distribution of AoD to be lower: the Furnished condition's peak in the highlighted area was higher than those of Empty condition, but the peak near zero was also higher than that of Empty one, as shown in Figure~\ref{fig:AoDResult}(A2). In the Small condition, the peak of highlighted area in the Furnished condition is higher than in the Empty condition, as shown through Figure~\ref{fig:AoDResult}(A3).

\subsection{Post VR Sickness Comparison}

A paired-sample t-test was conducted to compare the mean post-SSQ Total Score (TS) by the participants after conducting the Large $\times$ Empty (with gains) condition and the Large $\times$ Empty (no gain) condition. There was no significant difference in the mean post-SSQ TS scores for Large $\times$ Empty (with gains) (M = 27.58, SD = 24.87) and Large $\times$ Empty (no gain) (M = 24.27, SD = 23.95) conditions; t = 1.20, p = 0.249. A paired-sample t-test was also conducted to compare the mean post-SSQ TS score by the participants after conducting the Large $\times$ Furnished (with gains) condition and the Large $\times$ Furnished (no gain) condition. There was no significant difference in the mean post-SSQ TS scores for Large $\times$ Furnished (with gains) (M = 29.16, SD = 14.50) and Large $\times$ Empty (no gain) (M = 28.81, SD = 21.05) conditions; t = 0.076, p = 0.940. We thereby confirm that relative translation gains are not a significant factor for increases in VR simulator sickness.

\section{Discussion}

\subsection{Analysis}

We verify the six hypotheses based on our statistical analysis and AoD distribution graphs. Our results confirmed the first hypothesis (H1-1): The probabilities of ``larger'' responses at all relative translation gains in the Large condition were significantly lower than in the Medium condition. Our results also supported the second hypothesis (H1-2), as the probabilities of ``larger'' responses at all relative translation gains in the Large condition were significantly lower than in the Small condition. These results indicate that participants were less sensitive to the increase in their walking speed in the Large room than in the Medium and Small rooms. Thus, our results align with Kim et al.'s~\cite{kim2021adjusting} in that the relative translation gain thresholds increased to a greater degree when the virtual room size became larger. Considering that the shape of virtual spaces used in Kim et al.'s~\cite{kim2021adjusting} study and ours differed, the former rectangular and the latter square-shaped, we assert that the size of a VE is more relevant in determining the threshold range rather than the shape of the room or walking distance.

Furthermore, we were able to verify that the location of the virtual horizon affects the threshold range in Empty conditions by recording the user's AoD distribution, as shown in Figure~\ref{fig:AoDResult}. Figure~\ref{fig:AoDResult}(A1)-(A3) shows that in each Empty condition, users lay their gaze most frequently in the highlighted area, which indicates that they were aware of the virtual horizon while they were walking. Although the AoDs in three room size conditions were set incrementally, a linear relationship between the average AoD and estimated threshold values was not found. We assume that this is because the degrees of AoD within which user take notice of the virtual horizon continuously change while they are walking, and their height also affects the AoD distribution. As AoD values were affected by the height of participant's eye-gaze and the location of the virtual horizon, this implies that not only the size of a virtual room but also the height of the user should be considered as factors in determining the threshold range of relative translation gains.

On the other hand, the probabilities of ``larger'' responses at all relative translation gains in the Medium room were not significantly lower than in the Small condition, thus rejecting our third hypothesis (H1-3). We posit that this may be on account of the fact that the size difference between the Medium (8.2 m $\times$ 8.2 m) and Small (5.2 m $\times$ 5.2 m) condition was relatively smaller than the size differences between the Large (17 m $\times$ 17 m) and Medium (8.2 m $\times$ 8.2 m) or Large (17 m $\times$ 17 m) and Small (5.2 m $\times$ 5.2 m) condition. That the AoD distribution of Medium $\times$ Empty (purple line) and Small $\times$ Empty (red line) were similar as shown in Figure~\ref{fig:AoDResult}(A2),(A3) supports this assumption.

Our results further lead us to reject H2-1, H2-2, and H2-3, in which we postulated that the presence of virtual objects would lower the location to the virtual horizon, thereby decreasing both the AoD and the threshold range: Figure~\ref{fig:ARTResult} shows the probabilities of ``larger'' responses at all relative translation gains in the Furnished condition were significantly lower than in the Empty condition. This indicates that users were less sensitive to the change in their walking speed when objects were placed in the virtual space. Figure~\ref{fig:AoDResult}(A1) shows that the presence of furniture led to a shift in the AoD distribution for the Large rooms, meaning that the virtual horizon was perceived to be lower. Although we assumed that this would raise the level of user awareness regarding adjustments made to relative translation gains, statistical results were to the contrary.

Based on previous studies suggesting that visual cognitive load and attention influence the sense of vection in VEs~\cite{ruddle2011walking, trutoiu2008tricking}, we assert that this is because across the three furnished conditions, virtual objects were more dominantly perceived as distractions that drove the user's attention away from changes in their walking speed than as blockages placed to shorten the distance to the horizon and thereby reduce the perceived size of the virtual space. More recently, Williams et al.~\cite{williams2019estimation} showed that the rotational gain threshold increased when a distractor was present in the VE. We were also able to identify evidence from semi-structured interviews with the participants, in which they stated that the furniture in the virtual rooms had caught their attention. Some stated that in estimating their own speed, they relied on the speed at which the furniture moved towards them as they were walking as a point of reference. When asked how the overall experience during the task was, some commented that they kept thinking of the furniture around them, and one of them even ``walked imagining what would be behind the furniture."

This line of reasoning can also explain why the furnished condition was less effective in shifting the AoD distribution in the Medium and Small rooms towards the lower end than the Large room, as shown in Figure~\ref{fig:AoDResult}(A2),(A3): As the AoD value in perceiving the virtual horizon was already small in these size conditions compared to the Large room, objects in the room had a greater effect in distracting the users from the change in their walking speed and caused them to gaze upward more frequently. Table~\ref{tab:threshold} illustrates that the estimated threshold value $\alpha_T,upper$ in the Furnished conditions were more than 0.1 larger than those of the Empty conditions. This indicates that compared to the user's walking speed in the Empty conditions, that in the Furnished conditions could be increased as much as 10\% without them noticing the differences.

\subsection{Implications}

Relative translation gains allow users to move around in a more expansive virtual space than the real space they are in, all the while maintaining the alignment between the real and virtual space and thereby creating a more stable environment for spatial deformation. Our study shows that in the context of utilizing relative translation gains for these purposes, the effect of expanding or transforming the space can increase when the threshold range is shifted to a larger degree. Therefore, we propose two design considerations to enhance the effect of relative translation gains in expanding the size of the adjusted movable space in VEs.

First, the perceived movable space should be made larger than the adjusted movable space to obtain a wider range of relative translation gain thresholds in constructing virtual scenes. However, users may collide with the physical boundaries of the real space they are in when they attempt to walk towards the perceived movable space with this approach. Therefore, VR safety systems such as grid visualization, which is widely used for VR HMD-based playground setups\footnote{\url{https://support.oculus.com/guardian/?sf241303396=1}}, should be implemented in tandem. When a VR HMD application requires securing a broad walkable virtual space, developers may consider constructing the virtual space with expanded perceived movable space and limiting users' actual movable area through virtual object placement or other stimuli.

Second, placing objects in the virtual scene can further increase the range of relative translation gain thresholds and the size of the movable space, subsequently. Those who experienced the Furnished condition stated that while they used the speed at which the furniture in the space approached them to estimate their own speed, they were less sensitive to an increase in speed compared to the Empty condition. Based on these responses and the estimated threshold ranges in each condition, we conclude that even when the size of the perceived movable space remains the same, objects function as distractors and change the user's perception of their own walking speed in VE. Therefore, the developers should made informed choices regarding the RDW threshold depending on the existence of virtual objects in the virtual scene.

The implications derived from our study's findings can contribute to the generation of realistic and effective mutual spaces through RDW based on relative translation gains for MR remote collaboration. For example, organizing a mutual space in a large, furnished conference hall where the perceived movable space can be considerably larger than the adjusted movable space will be more beneficial than setting up an empty, nondescript virtual space configured by walls that limit the perceived movable space to match the adjusted movable space. As prior works have found that virtual objects are regarded as obstacles or distractors along with real ones in an AR scene as well~\cite{shin2021user, shin2019any}, leveraging the use of virtual objects in asymmetric collaboration will result in a mutual MR space more suitable for both the VR and AR side than other mutual space generation methods with empty spaces as the default setting~\cite{lehment2014creating}.

\subsection{Limitations}

Our study confirms that the threshold range for relative translation gains is affected by how users visually perceive a virtual space. However, there are some limitations to our approach. First, the effect of objects as distractors on the threshold range should be verified in detail, as a systematic, quantitative data-based analysis on the degree to which the objects attracted the user's attention was not given in our study. Although our study focused on the effects of the presence or absence of objects in virtual room, further studies on various configuration of virtual spaces through the placement of objects are needed. In addition, the effects of dynamic objects such as avatars or interactive objects should be also be further investigated.

Second, our study was conducted with users placed solely in a single space, which presents limits in applying its implications to multi-user scenarios in multiple heterogeneous spaces. When more than one user is simultaneously involved in RDW in a co-located mutual space, each with different relative translation gains applied according to the spatial configurations of their physical space, the presence of other users and their actions may alter the effect of space size and the existence of objects, as a previous study affirmed that the visual style and locomotion speed of another user's avatar influences the remote user's perception~\cite{choi2020effects}.

Lastly, the effects of individual differences~\cite{williams2019estimation, kim2021adjusting} on the cognitive threshold of relative translation gains should be explored. Although we set the gender ratio between the two groups to be the same and adjusted the average height between conditions to minimize the impact of these individual differences on the threshold range, further research considering these differences is needed to estimate personalized threshold ranges. By considering the effects of spatial configurations and personal traits together, the accuracy of relative translation gains and their threshold ranges can be further enhanced in generating a unified mutual space.

\section{Conclusion and Future Work}

In this study, we examined the effects of virtual room size and the presence of objects on the threshold range of relative translation gains in six spatial conditions, which were configured according to disparate Angles of Declination (AoDs) between the user's eye-gaze and forward-gaze. Relative translation gain thresholds are biased to be higher in large virtual rooms, as the actual AoD distribution shifts towards the higher end as the size of the room increases. The presence of virtual objects also increased the threshold range of relative translation gains, despite the fact that they had no apparent effect on shifting the AoD distribution. In the furnished rooms, objects functioned more as distractors that contributed to a relative insensitivity to the change in their walking speed, inducing users to become less sensitive be faster in the VE.

When creating a mutual space while maintaining the alignment of the real space and the virtual space, relative translation gains could be used as a technique to adjust the space of a VR user to the desired shape beyond existing methods that identify common areas from heterogeneous spaces~\cite{lehment2014creating, keshavarzi2020optimization}. Based on our findings, we conclude that constructing a virtual space where the perceived movable space is larger than the adjusted movable space, as well as placing objects in the space, can increase the space-deforming effect of relative translation gains. Applied in this way, relative translation gains can be used for adjusting heterogeneous virtual spaces to generate realistic MR mutual spaces that consider the configurations of both the physical and virtual space.

Our future study will investigate how varied conditions involving virtual objects influence users' attention during RDW, first by controlling the number of objects in a virtual scene. Furthermore, we will explore how the layout of virtual spaces configured by the placement of objects affects relative translation gain thresholds, as it was fixed to a single configuration in our current study. We will also analyze the effect of each type of distractor with quantitative data on the amount of time each distractor was gazed at. Finally, we will expand our study to various multi-user remote MR collaboration scenarios to verify the benefits of relative translation gains in such environments and how they influence user experiences.

\acknowledgments{
This work was supported by the National Research Council of Science \& Technology (NST) grant by the Korea government (MSIT) (No. CRC21$\sim$) and the National Research Foundation of Korea (NRF) grant funded by the Korea government (MSIT) (No. 2021R1A2C2011459).
}

\bibliographystyle{abbrv-doi}

\bibliography{template}
\end{document}